\documentclass[11pt]{article} 
\usepackage{epsfig} 
\begin{document} 

\title{Elastic response of [111]-tunneling impurities}
\author{Peter Nalbach, Orestis Terzidis\footnote{currently at SAP Labs
France, 505 route des Lucioles, F-06560 Valbonne} \\
  \it Institut f\"ur Theoretische Physik,
        Universit\"at Heidelberg,
        \\ Philosophenweg 19, 69120 Heidelberg, Germany
  \\ \\ \rm Karen A. Topp \\
  \it Laboratory of Atomic and Solid State Physics
  \\ Cornell University, Ithaca, New York, 14853
  \\ \\ \rm Alois W\"urger \\
  \it Universit\'e Bordeaux 1,
      CPMOH\footnote{Unit\'e Mixte de Recherche CNRS 5798}, 
       \\ 351 cours de la Lib\'eration, 33405 Talence, France}
\maketitle
 
\begin{abstract} 
We study the dynamic response of a [111] quantum impurity, such
as lithium or cyanide in alkali halides, with respect to an
external field coupling to the elastic quadrupole moment. Because
of the particular level structure of a eight-state system on a cubic
site, the elastic response function 
shows a biexponential relaxation feature and a van Vleck type
contribution with a resonance frequency that is twice the tunnel
frequency $\Delta/\hbar$. This basically differs from the dielectric
response that does not show relaxation. Moreover, we show that
the elastic response of a [111] impurity cannot be reduced to that
of a two-level system. In the experimental part, we report on recent
sound velocity and internal friction measurements on KCl doped with
cyanide at various concentrations. At low doping (45 ppm) we find the
dynamics of a single [111] impurity, whereas at higher concentrations
(4700 ppm) the elastic response rather indicates strongly correlated
defects. Our theoretical model provides a good description of the
temperature dependence of $\delta v/v$ and $Q^{-1}$ at low doping, 
in particular the relaxation peaks, the absolute values
of the amplitude, and the resonant contributions. From our fits we
obtain the value of the elastic deformation potential
$\gamma_t=0.192$~eV. 
\end{abstract}

\section{Introduction} 

The low-temperature properties of alkali halides may be significantly 
modified by the presence of substitutional impurities, such as Li, CN, 
or OH. Already a few ppm of these defects totally change the thermal
behavior and the elastic and dielectric response below Helium temperature. 

Regarding the polar molecules CN and OH, the point symmetry of the
impurity site gives rise to several equivalent orientations;
corresponding off-center positions arise for the small lithium ion.
Quantum tunneling between these states results in a ground state
splitting of about 1 Kelvin. Since the number of such tunneling states
exceeds, even at low concentration, the number of small-frequency phonon
modes of the host crystal, the impurities govern the low-temperature 
properties of the material. The cubic symmetry of fcc crystals favors 
8 defect positions in [111]-directions (CN and Li) or 6 positions in 
[100]-directions (OH in KCl). The resulting energy spectra have been 
discussed in detail by Gomez et al \cite{Gomez}, 
and agree well with the Schottky peak observed by Pohl and 
co-workers for various impurity systems  at low doping \cite{Nara}.

In this paper we are concerned with the elastic susceptibiltiy of a
[111]-impurity. Because of the cubic symmetry, such a tunneling system
behaves in many respects as an ensemble of three two-level systems with
energy splitting $\Delta$ \cite{Nara,Peter}. For example, the specific heat 
contribution
of a [111]-impurity is three times the Schottky peak of a two-state system.
A similar relation holds true for the dielectric response function, since
the dipolar transitions occur between adjacent levels only, as indicated by
the dashed lines in the actual energy spectrum shown in Fig. \ref{spectrum}.
This analogy has been used to describe lithium or cyanide defects in terms
of a two-state approximation \cite{Bau69,Kle84,Kra92,Ter96,Wur97}. 

The elastic response function, however, shows a more complex behavior,
resulting from the tensor character of the quadrupole operator $Q_{ij}$
and the structure of the energy spectrum shown in Fig. \ref{spectrum}.
According to the allowed quadrupolar
transitions indicated by solid arows in Fig. \ref{spectrum}, sound waves
or external stress perturb the impurity in two ways: First, they mix
states separated by twice the splitting, giving rise to a van Vleck type
susceptibility with resonance frequency $2\Delta/\hbar$. Second, they lift
the degeneracy of the states of a given triplet level, resulting in a
relaxation contribution to the response function. 

Early measurements in KCl:CN samples \cite{ByerSack} in the classical
regime ($k_{\rm B}T\gg\Delta$) showed that the change of sound velocity of a
$T_{2g}$-mode ($\hat{\bf e} = [110]$, $\hat{\bf k} = [001]$) varies
as expected like $1/T$ with temperature. Yet in addition, these experiments 
showed a change of sound velocity for an $E_g$-mode ($\hat{\bf e}
= [110]$, $\hat{\bf k} = [1$-$10]$) with an amplitude of about ten
times smaller than the $T_{2g}$ ones, whereas this mode should be
unaffected according to the simplest model of a symmetric [111]-tunneling
defect.

When measuring the sound velocity ( $T_{2g}$-mode ) of KCl:Li as a
function of temperature in the tunneling regime, H\"ubner et al. found
a hump at $k_{\rm B}T\approx \Delta$ \cite{Huebner}. 
They related this observation to relaxational transitions between
degenerate states, by calculating the static elastic response as the
second derivative of the free energy of the impurity ion. Their results
are valid in the low-frequency limit, where the external frequency
is smaller than the impurity relaxation rates.

The internal friction, however, cannot be obtained from a static
theory, and the low-frequency limit is not always justified for
the sound velocity. Starting from standard dynamic perturbation theory,
we develop a dynamic theory for the elastic response function of a
[111]-impurity coupled to a phonon heat bath. The main purpose of this
paper is to point out the peculiarities of the elastic response
function that arise from the degenerate states of the energy
spectrum shown in Fig. \ref{spectrum}. Moreover, we report sound
velocity and internal friction data of a $T_{2g}$-mode for KCl doped
with cyanide molecules, and discuss them in terms of our
[111]-impurity model with damping.

The outline of our paper is as follows.
In Sec. II we present the model and the basic equations 
for the elastic response function of a single impurity. The temperature 
dependence of the internal friction and sound velocity is evaluated in 
Sec. III. In Sec. IV we give some details of the experiments on KCl:CN, 
which are discussed in Sec. V. The final section contains a summary. 

\section{Theory}\label{theory}
\subsection{Quantum operators of a [111]-impurity} 

\begin{figure}[t]  
\epsfbox{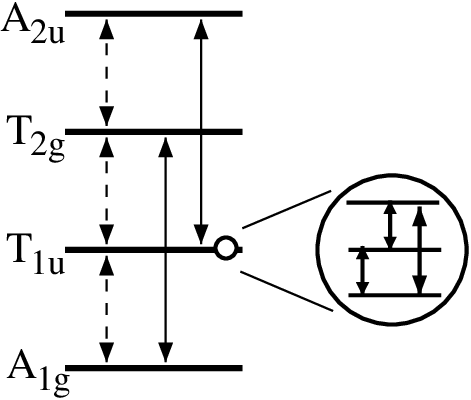}
\caption{\label{spectrum} Energy spectrum of a [111]-impurity with zero 
          asymmetry energy. Dashed arrows indicate the allowed dipolar 
          transitions, and full arrows the quadrupolar ones.}
\end{figure} 

We start by discussing the properties of the energy eigenstates and by 
presenting our notation based on pseudospin operators. As shown in Fig.
\ref{spectrum}, the energy spectrum contains four equidistant levels;
the upper and lower ones, A$_{\rm 1g}$ and A$_{\rm 2u}$, are single quantum
states, whereas the middle levels T$_{\rm 1u}$ and T$_{\rm 2g}$ are threefold
degenerate. This groundstate multiplet can be described as the direct product
of three two-level systems with energy splitting $\Delta$. 

Each energy eigenstate is a superposition of states $|{\bf r}\rangle$
localized at the eight impurity positions 
\begin{equation} \label{e2} 
{\bf r} = {1\over2}d(\sigma_z^1, \sigma_z^2, \sigma_z^3) 
\end{equation} 
with the effective two-state coordinates $\sigma_z^i$. For a lithium 
impurity, the off-center positions ${\bf r}$ form a cube of side-length 
$d$, whereas for cyanide impurities ${\bf r}$ indicates the orientation 
of the cigar-shaped polar molecule. Due to quantum tunneling along the 
axes $i=1,2,3$, the energy eigenstates factorize as 
$|\sigma_x^1\rangle|\sigma_x^2\rangle|\sigma_x^3\rangle$, where 
the variables $\sigma_x^i=\pm1$ label odd and even superpositions 
$|\sigma_x^i\rangle= 2^{-1/2}\,\{\,|r_i=d/2 \rangle + 
\sigma_x^i|r_i=-d/2\rangle\,\}\,$ along the $i$-axis. 

Accordingly, the quantum operators of the impurity may be written as
direct products of pseudospin operators $\sigma_\alpha^i$, where
$\alpha=x,y,z$ label the usual Pauli matrices and $\alpha=0$ the
identity operator. In terms of quantum states at $r_i=\pm d/2$ we have 
\begin{equation}\label{e4} 
\sigma_x^i=|d/2\rangle\langle -d/2| +|-d/2\rangle\langle d/2| 
\end{equation} \label{e6} 
\begin{equation}\sigma_z^i=|d/2\rangle\langle d/2| - |
-d/2\rangle\langle -d/2|
\end{equation} 
Thus a lithium or cyanide impurity is described as the product of 
three two-state variables. We adopt the shorthand notation for its 
quantum operators 
\begin{equation} \label{e8} 
A_{\alpha\beta\gamma} 
= \sigma_\alpha^1\otimes\sigma_\beta^2\otimes\sigma_\gamma^3 , 
\end{equation} 
where $i=1,2,3$ label the crystal axes and Greek indices 
$\alpha=0,x,y,z$ the pseudospin operators. 
When denoting by ${1\over2}\Delta$ the tunneling amplitude between 
adjacent impurity positions, we obtain the Hamiltonian  
\begin{equation}\label{Hsys} 
H_S = -{\Delta \over 2}\left( A_{x00} + A_{0x0} + A_{00x} \right). 
\end{equation} 
Its eigenstates may be labelled according to the eigenvalues $\pm1$ of 
$\sigma_x^i$; for example, the ground state A$_{\rm 1g}$ reads as 
$|$+++$\rangle$, and the states of the triplet level T$_{\rm 1u}$ as 
$|-$++$\rangle$, $|$+$-$+$\rangle$, and $|$++$-\rangle$. 
According to the above discussion, the statistical operator 
$\rho={\rm e}^{-\beta H_S}/{\rm tr}({\rm e}^{-\beta H_S})$ factorizes as 
\begin{equation}\label{rho} 
\rho = \rho^1 \otimes \rho^2 \otimes \rho^3 , 
\hskip1cm 
\rho^i = \frac{1}{2} 
\left( 1 + \sigma_x^i\tanh(\Delta/2k_{\rm B}T) \right) . 
\end{equation} 
(In the limit of zero temperature, the even state $\sigma_x^i=1$ is 
occupied with probability 1.) From the statistical operator it is 
evident that the specific heat anomaly due to $N$ such defects is 
identical to the Schottky peak of $3N$ two-level systems with the 
Hamiltonian ${1\over2}\Delta\sigma_x$. As to the dynamic properties, 
a similar relation holds true for the dielectric susceptibility. 
This is not surprising, since dipolar transitions involve a single 
coordinate $r_i$ and occur between adjacent levels with splitting 
$\Delta$ only; cf. Fig. \ref{spectrum}. 

A different situation, however, is encountered when considering the 
elastic response: The quadrupole operator 
\begin{equation} \label{e10} 
Q_{ij} = r_ir_j ( 1- \delta_{ij}) 
\end{equation} 
involves two coordinates, and hence causes transitions between 
energy eigenstates that differ in two labels $\sigma_x^i$. Thus 
elastic perturbations induce two types of transitions as indicated 
by the solid arrows in Fig. \ref{spectrum}. 

\subsection{Phonon coupling}

Noting (\ref{e2}) and $A_{z00}A_{0z0}=A_{zz0}$, we may write the 
quadrupole operators as $Q_{12}=(d/2)^2A_{zz0}$, etc. When absorbing 
the factor $(d/2)^2$ in the coupling constant $\gamma$, we obtain  
the impurity-phonon interaction \cite{Wur97} 
\begin{equation}\label{e12} 
H_{SB} = 2 \sum_\alpha \gamma_\alpha \left(A_{zz0}\,
\epsilon_{12}^{\alpha} + A_{z0z}\, \epsilon_{13}^{\alpha} 
+ A_{0zz}\, \epsilon_{23}^{\alpha} \right) . 
\end{equation} 
When evaluating the strain tensor at the impurity site ${\bf R}$
and taking the limit of long wavelenths, $kd\ll1$, we find
\begin{equation}\label{e14} 
\epsilon_{ij}^{\alpha}= \frac{i}{2} 
\sum_{\bf k} \sqrt{\frac{\hbar}{2m\omega_{{\bf k}a}}} {\rm e}^{i{\bf
k}\bf{R}} 
\left(e_i^{\alpha}({\bf k})k_j+e_j^{\alpha}({\bf k})k_i\right) 
\left(b_{{\bf k},\alpha}+b_{-{\bf k},\alpha}^{\dag}\right). 
\end{equation} 
Here $\bf{k}$ and ${\bf e}^{\alpha}$ are wave and polarisation vectors,
and $\alpha$ labels the longitudinal and transverse phonon branches. 

\subsection{Asymmetry energy}\label{asymmetry}

Up to now we have considered an impurity in a potential that reflects the 
cubic site symmetry. The strong dipolar and elastic interactions of nearby 
impurities, however, break that symmetry and give rise to an effective 
asymmetry energy, 
\begin{equation}\label{asymmetrie} V = -\frac{v_1}{2} \,A_{z00}
-\frac{v_2}{2} \,A_{0z0}  
-\frac{v_3}{2} \,A_{00z} , 
\end{equation} 
where the $v_i$ are random quantities with zero mean. We assume a Gaussian 
distribution with width $\sigma$: 
\begin{equation} P(\vec{v})=\prod_iP_i(v_i) \qquad\mbox{with}\qquad
P_i(v_i) = \frac{1}{\sqrt{2\pi}\sigma}\exp\left(-\frac1
2\,\frac{v_i^2}{\sigma^2}\right)  
\end{equation} 
In principle, the $v_i$ are dynamic variables closely related to the
position  operators of nearby impurities. In the case of weak
interaction, $v_i\ll\Delta$,  
however, they may be treated as static random fields. Eq. (\ref{asymmetrie}) 
accounts for random fields arising from dipolar interactions; elastic
coupling would lead to a potential that involves quadrupole operators such
as $A_{zz0}$. 

\section{Elastic response function}

The lattice vibrations of the host crystal act on the tunneling 
impurity as a heat bath at temperature $T$. Upon a perturbation 
by an external strain field, the impurity regains the thermal 
equilibrium through absorption or emission of resonant phonons. 
Since the impurity-phonon coupling is weak, the resulting damping 
rates can be evaluated in second-order perturbation theory. 

\subsection{Dynamic perturbation theory}

The linear response of the impurity to a time-dependent external 
strain field is described by the commutator correlation function 
$\langle [Q_{ij}(t),Q_{ij}]\rangle$. For convenience, we use 
dimenensionless quadrupole operators, e.g. $A_{zz0}=(2/d)Q_{12}$
with $\langle A_{zz0}^2\rangle=1$; all relevant response functions
can be expressed in terms of
\begin{equation}\label{e15} 
\chi(t) = \langle [A_{zz0}(t),A_{zz0}]\rangle 
\end{equation} 
It turns out convenient to calculate first the correlation matrix 
\begin{equation}\label{e16} 
C_{\alpha\beta\gamma;\delta\epsilon\zeta}(t) 
= (A_{\alpha\beta\gamma}(t)|A_{\delta\epsilon\zeta}), 
\end{equation} 
where we have defined the symmetrized correlation function 
\begin{equation}\label{e18} 
(A|B) = {1\over2}\langle AB + BA \rangle. 
\end{equation} 
Our evaluation of the correlation matrix is based on the 
Mori-Zwanzig projection formalism; the resulting memory kernel 
is calculated to second order in the impurity-phonon coupling 
\cite{Mori}. Thus we obtain a matrix equation for the Laplace 
transform of (\ref{e16}), 
\begin{equation} \label{e20} 
C(z) = \frac{-1}{z-\Omega-K(z)} M, 
\end{equation} 
with the usual definitions of the static correlations 
\begin{equation} \label{e22} 
M_{mn}=(A_m|A_n)=(\eta^{-1})_{mn} , 
\end{equation} 
the frequency matrix 
  \begin{equation} \label{e24} 
  \Omega_{mn} = (A_m|{\cal L} A_p)\eta_{pn} , 
  \end{equation} 
and the memory kernel 
\begin{equation} \label{e26} 
K_{mn}(z) = (\dot A_m| (z-{\cal L}_0)^{-1} |\dot A_p) \eta_{pn} . 
\end{equation} 
For convenience, we have replaced the triple index pair of (\ref{e8}) 
with $m,n$, and we have used the shorthand notations $\dot A_i= 
{\rm i}[H_{SB},A_i]$ and $\hbar{\cal L}_0*=[H_S+H_B,*]$, where
$H_B=\sum\hbar\omega_k  
b_k^\dagger b_k$. The thermal average $\langle...\rangle$ is with 
respect to $\rho$ and the equilibrium distribution of the phonons. 

Solving the impurity dynamics now amounts to calculating the 
eigenvalues und residues of the resolvent $[z-\Omega-K]^{-1}$. 
Since there are $4^3$ operators $A_{\alpha\beta\gamma}$, the matrix 
equation involves a 64-dimensional space. The matrices $C$, $M$, 
$\Omega$, and $K$ being block-diagonal, the actual problem simplifies 
significantly. 

For symmetry reasons, the time evolution of the three operators 
$A_{zz0}$, $A_{z0z}$, and $A_{0zz}$ is the same. Therefore it is 
sufficient to evaluate the dynamics in the invariant subspace 
containing one of them, e.g., 
\begin{equation} \label{e27}\begin{array}{l} 
A_1 = A_{zz0},\; A_2 = A_{zy0},\; A_3 = A_{yz0},\; A_4 = A_{yy0}, 
\\ 
A_5 = A_{zzx},\; A_6 = A_{zyx},\; A_7 = A_{yzx},\; A_8 = A_{yyx}. 
\end{array} \end{equation}
In this invariant subspace, $\Omega$, $M$, and $K$ are represented 
by $8\times8$-matrices that are easily calculated according to 
(\ref{e22}--\ref{e26}). Note that the commutation relations for 
the operators $A_i$ can be traced back to those for Pauli matrices. 
It can be shown that $\Omega$ and $M$ can be diagonalized 
simultaneously; in the present example it is straightforward to find 
the corresponding unitary transformation $U$ and to obtain 
\begin{equation} \label{e28} 
\widetilde{\Omega}=U^{\dagger}\Omega U 
=diag( 0, 0, 0,0,-2\Delta, -2\Delta, 2\Delta, 2\Delta)/\hbar 
\end{equation} 
\begin{equation} \label{e30} 
\widetilde{M}=U^{\dagger} M U =diag(m_1,m_1,m_2,m_2,m_3,m_4,m_3,m_4) 
\end{equation} 
with 
\begin{displaymath} \begin{array}{l} m_1 = (1-t)\,(1-t^2) \\ 
m_2 = (1+t)\,(1-t^2) \\ 
m_3 = (1+t)\,(1+t^2) \\ 
m_4 = (1-t)\,(1+t^2) 
\label{e32} 
\end{array}\end{displaymath} 
where we have used $\beta=1/k_{\rm B}T$ and $t=\tanh(\beta \Delta/2)$. 

Besides the fourfold degenerate zero frequency, there are two doubly 
degenerate finite frequencies $\pm2\Delta/\hbar$ in Liouville space. 
While $\Omega$ and $M$ could be diagonalized simultaneously, this 
is not possible for the memory matrix $K$. It turns out, however, that 
its transform $\widetilde K$ is diagonal in each of the three degenerate 
subspaces of $\widetilde \Omega$, with bare frequencies $0,
\pm2\Delta/\hbar$;  
the finite off-diagonal entries connect, e.g., the subspace of zero 
frequency eigenvalue with that of eigenvalue $2\Delta/\hbar$. Because 
of $\widetilde K\ll\Delta/\hbar$, we may neglect these off-diagonal entries 
of $\widetilde K$; they would result in very small corrections to the 
frequencies and residues of the order of $(\hbar\widetilde K/\Delta)^2$. 
In physical terms, the off-diagonal parts of $\widetilde K$ in the 
degenerate subspaces vanish, since they involve the phonon 
density of states at zero frequency; in a three-dimensional 
body, however, the density of phonon modes vanishes in the limit 
$\omega_k\to0$.

Applying the usual Markov approximation, we evaluate the memory 
functions $\widetilde K(z)$ at the corresponding bare frequencies 
$z=0, \pm2\Delta/\hbar$. It turns out convenient to separate real 
and imaginary parts according to $\widetilde K=\widetilde K'+{\rm
i}\widetilde K''$.  
Whereas the entries in the subspace of zero eigenvalue are purely 
imaginary, those belonging to the finite frequencies 
$\pm2\Delta/\hbar$ are complex numbers. Their real parts 
$\widetilde K'(\pm2\Delta/\hbar)$ would induce a shift of the resonance 
frequencies from the bare values $\pm2\Delta/\hbar$ to slightly 
smaller ones. This shift being very small, we discard it and retain 
the imaginary, or dissipative, part $\widetilde K''$ only. The latter 
form a diagonal matrix $\widetilde\Gamma=\Im \widetilde K''(z_0)$ with 
$z_0=0, \pm2\Delta/\hbar$. 

With these approximations for the memory kernel, $\widetilde\Omega$, 
$\widetilde M$, and $\widetilde\Gamma$ are diagonal; the correlation
matrix thus factorizes, 
\begin{equation} \label{e34} 
\widetilde{C}_{jj}(z) = 
\frac{-\widetilde M_{jj}}{z-\widetilde{\Omega}_{jj}+{\rm
i}\widetilde\Gamma_{jj}} ,  
\end{equation} 
where $j=1,...,8$ labels the eight eigenvalues given in 
(\ref{e28},\ref{e30}) and those of the damping matrix $\widetilde\Gamma$. 
Since we are interested in the correlation function of the operator 
$A_{zz0}$, we have to calculate the element $C_{11}$ of the correlation 
matrix in the original basis, $C=U\widetilde C U^\dagger$. The
transformation $U$ being unitary, we have 
\begin{equation} \label{e36} 
{C}_{11}(z) = \sum_{j}|U_{1j}|^2 \widetilde C_{jj}(z) , 
\end{equation} 
where the vector of coefficients 
\begin{equation} \label{e38} 
\left(|U_{1j}|^2\right)_j = (0, 1/4, 1/4, 0, 1/8, 1/8, 1/8, 1/8) 
\end{equation} 
satisfies the condition $\sum_j|U_{1j}|^2=1$.

The damping rates $\widetilde\Gamma_{jj}$ are calculated to second-order 
in terms of the impurity-phonon coupling potential (\ref{e12}). Each 
rate is given as the convolution of an uncoupled correlation spectrum 
$C_{ii}^0(\omega)$ and the dissipation spectrum (\ref{app4}). The 
derivation is straightforward; we merely give the result 
\begin{equation} \label{e40} 
\widetilde{\Gamma} =
diag(\Gamma_1,\Gamma_1,\Gamma_2,\Gamma_2,\Gamma_3,\Gamma_4,\Gamma_3,\Gamma_4), 
\end{equation} 
with 
\begin{equation}\label{gamma} \begin{array}{l} 
\Gamma_1= \Gamma_0 (1+n(2\Delta)) \\ 
\Gamma_2= \Gamma_0 n(2\Delta) \\ 
\Gamma_3= \Gamma_0 (1+4n(2\Delta)) \\ 
\Gamma_4= \Gamma_0 (3+4n(2\Delta)) . \end{array}\end{equation} 
Here we have used the Bose occupation factor 
\begin{equation}\label{e44} 
n(2\Delta)=\frac{1}{{\rm e}^{\beta 2\Delta}-1} 
\end{equation} 
and the coupled phonon Green function (\ref{app4}) evaluated at the 
frequency $2\Delta/\hbar$, 
\begin{equation} \label{e46} 
\Gamma_0 = 2\Gamma''(2\Delta/\hbar) = \pi
\sum_{\alpha}f_\alpha{\gamma_\alpha^2  
\over v_\alpha^5} 
{(2\Delta)^3 \over 2\pi^2 \rho \hbar^4}. 
\end{equation} 
The geometric factor $f_\alpha$ arises from the different weight for 
longitudinal and transverse modes in the coupling matrix elements.

Now we have an explicit expression for the frequency dependent 
function (\ref{e36}). After inverse Laplace transformation we obtain 
the time-dependent correlation function of the operator $A_{zz0}$, 
\begin{equation}\label{e50} 
C_{11}(t) = \frac{1}{4} \left(m_1 {\rm e}^{-\Gamma_1 t} 
+ m_2 {\rm e}^{-\Gamma_2 t} 
+ m_3 \cos(2\Delta t+\delta_3) {\rm e}^{-\Gamma_3 t} 
+ m_4 \cos(2\Delta t+\delta_4) {\rm e}^{-\Gamma_4 t} \right) , 
\end{equation} 
which constitutes the basic theoretical result of this paper. 

As the most salient features we note the presence of two relaxation 
features with different relaxation rates $\Gamma_1$ and $\Gamma_2$.
The remaining terms (the socalled resonant terms) show 
oscillations with twice the tunneling frequency $\Delta/\hbar$ and 
different damping rates, with $\tan\delta_i=\hbar\Gamma_i/2\Delta$. The 
amplitudes $m_i$ vary significantly with temperature; in the 
high-temperature limit all of them tend towards unity, $m_i\to1$, 
whereas for $T\to0$ one finds $m_3=4$ and $m_1=m_2=m_4=0$. Thus the 
relaxation contributions disappear at zero temperature.

The spectra of the response function (\ref{e15}) and the correlation 
function (\ref{e50}) are related through the fluctuation-dissipation 
theorem, 
\begin{equation} \label{e52} 
\chi''(\omega) = (2/\hbar) \tanh(\beta \hbar\omega /2) C_{11}''(\omega) . 
\end{equation} 
With the usual approximations for the hyperbolic tangent function we 
thus obtain 
\begin{equation} \label{e54} 
\chi''(\omega) = \beta\hbar\omega \left[L_1(0) + L_2(0) \right] 
+ \tanh(\beta\Delta) 
\left[L_3(2\Delta) - L_3(-2\Delta) 
+ L_4(2\Delta) - L_4(-2\Delta) \right] , 
\end{equation} 
with the weighted Lorentzian 
\begin{equation} \label{e56} 
L_i(E) = {m_i\over4} 
{\hbar\Gamma_i \over (\hbar\omega-E)^2 + \hbar^2\Gamma_i^2}. 
\end{equation} 
Clearly, the spectrum of the response function $\chi''(\omega)$ 
contains the same physics as the correlation function $C_{11}$; 
hence the discussion below (\ref{e50}) applies equally well to 
(\ref{e54}). 

The tunneling impurities can be probed by an external ultrasonic wave 
with propagation vector ${\bf k}$ and polarisation vector 
${\bf e}^{\alpha}$. The attenuation of the latter is given by the 
internal friction 
\begin{equation} \label{e58} 
Q_\alpha^{-1} = n\, h_{{\bf k}\alpha}
{\gamma_\alpha^2\over\varrho v_\alpha^2} 
\chi''(\omega), 
\end{equation}
whereas the variation of the sound velocity 
\begin{equation} \label{e60} 
{\delta v_\alpha \over v_\alpha} = 
n\, h_{{\bf k}\alpha}
  {\gamma_\alpha^2\over2\varrho v_\alpha^2} 
\chi'(\omega), 
\end{equation} 
depends on the real part of the impurity dynamic susceptibility $\chi'$. 
Here $n$ is the number density of the impurities; the geometric factor 
$h_{{\bf k}\alpha}$, as defined in (\ref{app16}), accounts for the 
orientation of ${\bf k}$ and ${\bf e}^{\alpha}$ with respect to the 
crystal axes. 

Since the tunneling frequency $\Delta/\hbar$ is of the order of 20 GHz, 
the second term of (\ref{e54}) is immaterial at acoustic frequencies; 
and the internal friction involves the relaxational part only, 
\begin{equation} \label{e62} 
Q_\alpha^{-1} = n\, h_{{\bf k}\alpha} 
{\gamma_\alpha^2\over\varrho v_\alpha^2} 
{1\over k_{\rm B}T} 
\left[m_1{\omega\Gamma_1 \over \omega^2 + \Gamma_1^2} 
+ m_2{\omega\Gamma_2 \over \omega^2 + \Gamma_2^2} \right] . 
\end{equation} 
Note that both amplitudes are exponentially small at low temperatures, 
$m_1+m_2 = 2 *$ $\mbox{sech}(\beta\Delta/2)^2$, though $m_1$ vanishes
faster than  
$m_2$. On the other hand, $\Gamma_1$ tends towards the constant $\Gamma_0$, 
whereas $\Gamma_2$ disappears. Thus in the intemediate range 
$k_{\rm B}T\approx\Delta$ both contributions may be relevant. 

Because of the different temperature dependence of the two rates and 
the corresponding amplitudes, the variation of $Q^{-1}$ with frequency 
changes very much with the ratio $\omega/\Gamma_0$. 
The origin of the terms in (\ref{e62}) is quite obvious in view of 
the level scheme of Fig. \ref{spectrum}. The contribution 
involving $\Gamma_2$ stems from relaxation between the states of the 
level T$_{\rm 1u}$; the required intermediate transition to the 
top level A$_{\rm 2u}$ accounts for the temperature factor of $\Gamma_2$. 
On the other hand, the term with $\Gamma_1$ results from the upper triple 
level T$_{\rm 2g}$; relaxation occurs through the intermediate state 
A$_{\rm 1g}$; phonon emission in this downscattering process gives 
rise to the temperature factor of $\Gamma_1$. The rates of the resonant 
contributions in (\ref{e54}) can be discussed in a similar fashion. 

Finally we note that the relaxation rates $\Gamma_1$ and $\Gamma_2$ satisfy 
the condition $\Gamma_1={\rm e}^{\beta2\Delta}\Gamma_2$. They are
related to the phase decoherence rates $\Gamma_3$ and $\Gamma_4$ through 
$\Gamma_1+\Gamma_2=(1/4)(\Gamma_3+\Gamma_4)$; thus the transverse
rates are larger  
than the longitudinal ones, contrary to the well-known ratio for the 
two-level system, $1/T_1=2/T_2$. 

\subsection{Effects of asymmetry}

In order to obtain a realistic model for tunnel impurities, we have 
to account for the random asymmetry fields (\ref{asymmetrie}). At 
sufficiently low concentrations, $c<100$~ppm or 
$n<10^{19}$~cm$^{-3}$, typical values of $v_i$ are significantly 
smaller than the tunnel energy $\Delta$. Thus we may use a perturbation 
expansion in powers of the small parameter $v_i/\Delta$. 

Each pole of the dynamic susceptibility (\ref{e54}) is characterized 
by an amplitude $m_i$, a damping rate $\Gamma_i$, and an oscillation 
frequency that takes the values $\pm2\Delta$ or $0$. It turns out 
that the lowest-order corrections to $m_i$ and $\Gamma_i$ are quadratic 
in $v_i/\Delta$ and thus are of little significance. Similarly, the 
finite poles at $\pm2\Delta$ are hardly affected by the small asymmtry 
energies $v_i$. 

A significant change occurs, however, in the zero-frequency poles. 
When evaluating (\ref{e20}) with both (\ref{Hsys}) and (\ref{asymmetrie}) 
and calculating the lowest-order corrections in the diagonal representation 
(\ref{e34}), we obtain the set of frequencies 
\begin{equation} 
\widetilde{\Omega}\approx diag(\eta,\eta,-\eta,-\eta,-2\Delta,
-2\Delta, 2\Delta, 2\Delta) 
\end{equation} 
with 
\begin{equation} 
\eta=1/2\,(v_1^2-v_2^2)/\Delta. 
\end{equation} 
Accordingly, the relaxation contributions $L_1(0)$ and $L_2(0)$ in
(\ref{e54}) have to be replaced by ${1\over2}(L_i(\eta)+L_i(-\eta))$,
with $i=1,2$.  

Since experimentally frequencies are much smaller than the tunnel frequency 
$\Delta/\hbar$, the imaginary part of the susceptibility is determined by the 
low-frequency poles $\pm\eta$. When denoting the average with respect to the 
random fields $v_i$ by a bar, we have 
\begin{equation}\label{plo2} 
\overline{\chi''(\omega)} = {1\over2}\beta\omega 
\left( \overline{L_1(\eta)} + \overline{L_1(-\eta)} 
+ \overline{L_2(\eta)} + \overline{L_2(-\eta)} \right). 
\end{equation} 
When inserting the inverse Fourier transform of the Lorentzians, the 
integral over $v_i$ involves a Gaussian with complex width parameter 
$(1/2\sigma^2)\pm i t/\hbar\Delta$. Performing the Gaussian integrals 
we find 
\begin{equation}\label{chiasy2} 
\overline{\chi''(\omega)} = \beta\frac{\omega\Delta}{4\sigma^2} 
\int_{0}^{\infty}\mbox{d}t 
\frac{\cos(\widetilde{\omega}t) }{\sqrt{1+t^2}} 
\left(m_1e^{-\widetilde{\Gamma_1}t} + m_2 e^{-\widetilde{\Gamma_2}t}\right), 
\end{equation} 
with $\widetilde{\omega}=\omega\Delta/\sigma^2$ and 
$\widetilde{\Gamma}=\Gamma\Delta/\sigma^2$. We recall that we have used 
$k_BT,\Delta\gg\hbar\omega,\hbar\Gamma$. 

Now we turn to the real part that is obtained from (\ref{e52}) through 
the Kramers-Kronig relation. Proceeding as above we insert the inverse 
Fourier tranform; resorting to the usual approximations for the 
temperature-dependent factors and assuming $\beta\eta/2\ll 1$, we 
obtain the susceptibility 
\begin{equation}\label{plo1} 
\chi'(\omega) = \frac{m_3 + m_4}{4}\,\frac1 \Delta \tanh(\beta\Delta) 
\,+\,\beta\sum_{i=1,2}\sum_\pm\,\frac{m_i}{4} \frac{\eta(\eta\pm\omega) 
+ \Gamma^2_i}{ (\eta\pm\omega)^2 + \Gamma_i^2}. 
\end{equation} 
Averaging over the distribution of asymmetries yields 
\begin{equation}\label{chiasy1}
\overline{\chi'(\omega)}\approx\frac{m_3 + m_4}{4}\,\frac1 \Delta
\tanh(\beta\Delta)\,+\,\sum_{l=1,2}\,\frac{m_l}{4}\beta
\left\{1-\frac{\omega\Delta}{\sigma^2} \int_{0}^{\infty}\mbox{d}t
\frac{\sin(\widetilde{\omega}t) {\rm
e}^{-\widetilde{\Gamma_l}t}}{\sqrt{1+t^2}} \right\} 
\end{equation} 

\subsection{Comparison with a two-level tunneling system} 

When investigating the dynamic properties of substitutional impurities, 
various authors have used a two-state approximation. For this reason we 
briefly discuss the response function of a two-level system (TLS) and 
compare it with the present results for the eight-state system. 
With the notation introduced in Sect. 2, the Hamiltonian of a symmetric 
TLS reads as 
\begin{equation} \label{e64} 
H={1\over2}\Delta\sigma_x 
+ \sigma_z \sum_{\alpha,i,j}\gamma_\alpha \epsilon_{ij}^\alpha 
+ H_{\rm phonon} , 
\end{equation} 
which is nothing but the one-dimensional version of (\ref{Hsys}) and 
(\ref{e12}). When evaluating the dissipative two-state dynamics one finds 
that there is no relaxation contribution to the response function of a 
TLS; its susceptibility is of purely "resonant" character and similar to 
the second term of (\ref{e54}). 

Thus the two-state approximation misses the observed relaxation peak 
in the dynamic response. In order to mend this failure, an asymmetry 
energy has been added to (\ref{e64}), yielding $H={1\over2}\Delta\sigma_x 
+{1\over2}\epsilon\sigma_z$, with a two-level splitting $E=\sqrt{\Delta^2 
+\epsilon^2}$. The spectrum of the dynamic susceptibility then reads as 
\begin{equation}\label{qq} 
\chi''_{zz} = \frac{\Delta^2}{E^2} 
\tanh(\beta E/2 )\sum_\pm\frac{\pm\Gamma}{(\omega\mp E/\hbar)^2+\Gamma^2}\, 
+ \frac{\epsilon^2}{E^2} \mbox{sech}^2(\beta E/2) 
\frac{\beta\hbar\omega\cdot(2\Gamma)}{\omega^2+(2\Gamma)^2}. 
\end{equation} 
The amplitude of the relaxation feature is proportional to $\epsilon^2/E^2$ 
and thus vanishes in the limit of zero asymmetry energy. 
This feature distinguishes the actual eight-state system from the 
two-state approximation. The former contains a strong relaxation feature, 
even at zero asymmetry energy. The two-state approximation, however, fails 
to account for relaxation in the symmetric case $\epsilon=0$. 

\section{Experimental detail and results} 

The crystals measured here were seed pulled from the melt 
at the crystal growth facility of the Cornell Center for Materials 
Research. The starting powder of KCl was Merck 
Industries "Superpure" grade with nominal impurity levels of less than 
1 ppm. (A check of the OH$^-$ level in our crystal of "pure" KCl, via 
UV absorption, confirmed a concentration of 0.5 ppm.) The cyanide 
added to the melt was taken from a seed-pulled single crystal of 
KCN for which the starting powder had been vacuum baked to remove 
H$_2$O. 

The composition of the final crystal was analyzed via infra-red absorption 
($\sim$ 2100 cm$^{-1}$) to an estimated accuracy of 10\% relative error, 
by comparing the area of the absorbance peak to a standard measured 
for us by Fritz L\"{u}ty at the University of Utah 
where he measured the CN$^-$ vibrational spectra at liquid nitrogen 
temperature and calculated the concentration based on the known 
oscillator strength of CN$^-$\cite{luety74}. 

\begin{figure}
\centering{\epsfig{file=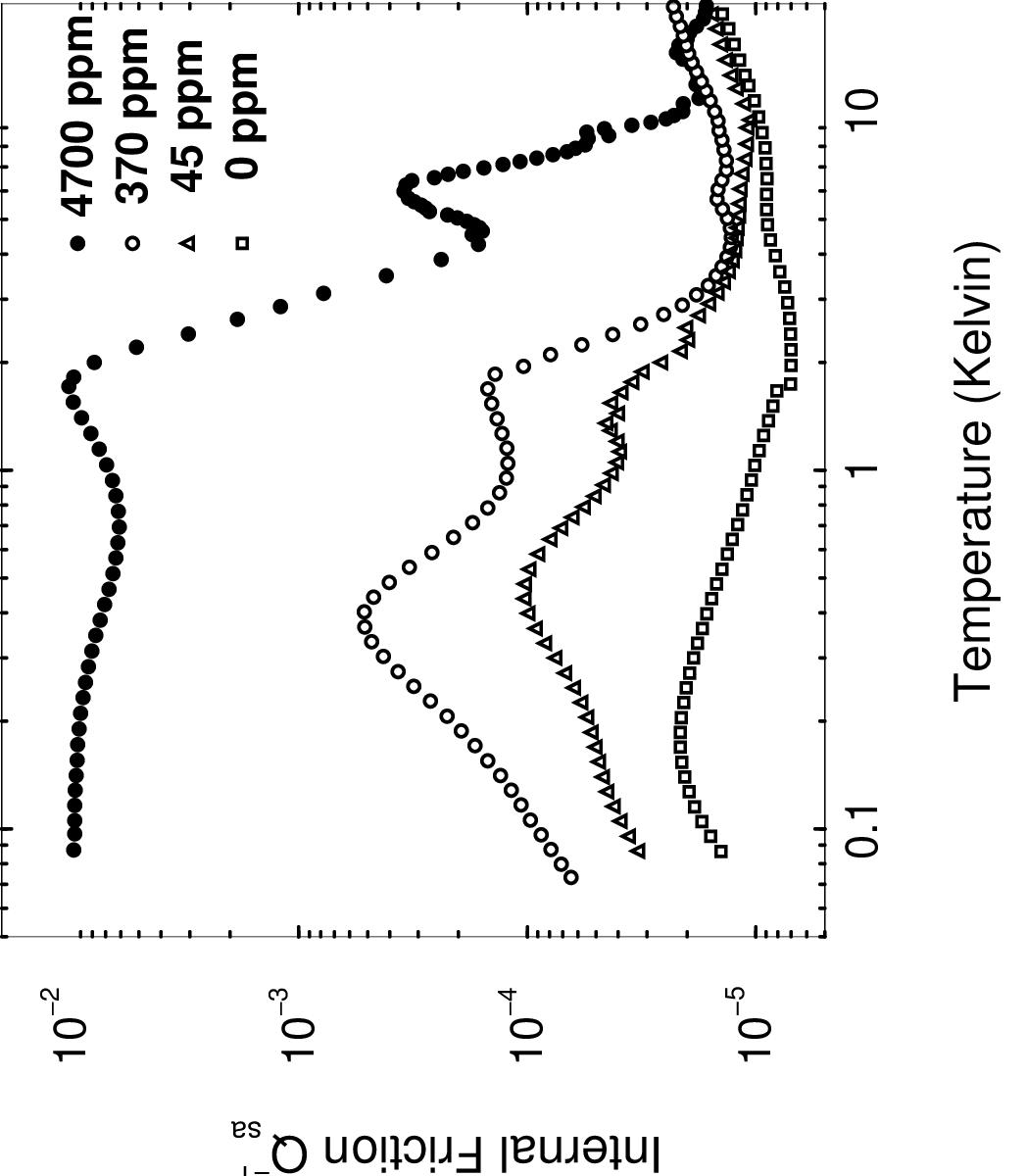,width=8cm,angle=270}} 
\caption{\label{Q_Daten} The internal friction data of four KCl samples 
with different concentrations of CN$^-$ dopants are plotted versus the 
temperature.} 
\end{figure}   

The internal friction sample was cleaved directly from the IR 
absorption sample to ensure a known CN$^-$ concentration. 
The internal friction was measured with a composite torsional oscillator 
as described in Ref.~\cite{cahillvanc}. In this method, a 90~kHz 
quartz transducer (2.5~mm diameter) and the sample form a composite 
torsion bar. The quartz end is attached to a thin Be-Cu pedestal 
\cite{cahillvanc} by an approximately 25~mg drop of Stycast 2850FT 
epoxy, and since the KCl crystals are quite fragile, a 0.25~mm indium 
foil was epoxied between the crystal and transducer to "cushion" the 
difference in thermal contraction rates of the two components upon 
being cooled. The sample length is tuned to be one half of a shear 
wavelength, so that the composite oscillator has a resonance frequency 
at room temperature within 1\% of the bare quartz crystal resonance. 
This adjustment ensures that the epoxy and indium joint between the 
quartz and sample has almost zero strain, and therefore contributes 
minimally to the observed internal friction. (The epoxy and 
epoxy/indium junctions produce a background contribution to the 
internal friction of less than 10$^{-5}$ at low temperatures.) 

The oscillator is driven by a set of electrodes which form a quadrupole 
configuration around the transducer and which simultaneously drive and 
detect its motion. The internal friction of the sample 
$Q^{-1}_{sa}$ is determined from the quality factor 
$Q_{comp}$ of the composite oscillator resonance by 
\begin{equation} 
Q^{-1}_{sa}=\left[{I_{sa}+(1+\alpha)I_{tr} 
\over I_{sa}}\right] Q^{-1}_{comp} 
\end{equation} 
where $I_{tr}$ and $I_{sa}$ are the moments-of-inertia 
of the transducer and sample respectively, and $\alpha$ is a 
correction for the attachment of the transducer to the thin Be-Cu 
pedestal, $\alpha \approx 0.06$\cite{cahillvanc}. The relative change 
in speed of sound can be found from the change in resonance frequency 
of the composite oscillator 
\begin{equation} 
\frac{\Delta v}{v}_{sa}=\left[{I_{sa}+(1+\alpha)I_{tr} 
\over I_{sa}}\right] \frac{\Delta f}{f} , 
\end{equation} 
where $\Delta f/f$ is determined relative to an arbitrary 
reference frequency usually at the coldest temperature of the run. 
Details on evaluating internal friction from a torsinal oscillator
can be found in Ref.~\cite{topp99}.
The measurements 
below 1.5~K were made in a dilution refrigerator, and those from 1.5 
to 300~K in an insertable $^4$He cryostat \cite{Swartz}. 

We have investigated a $T_{2g}$-mode ( $\hat{\bf e} = [110]$,
$\hat{\bf k} = [001]$ ) in three samples doped with 45, 370 and 4700 
ppm CN$^-$, and a pure sample; the latter provides the background 
due to the host crystal and thus permits us to determine the impurity 
contribution to the elastic response function. 

\begin{figure}
\centering{\epsfig{file=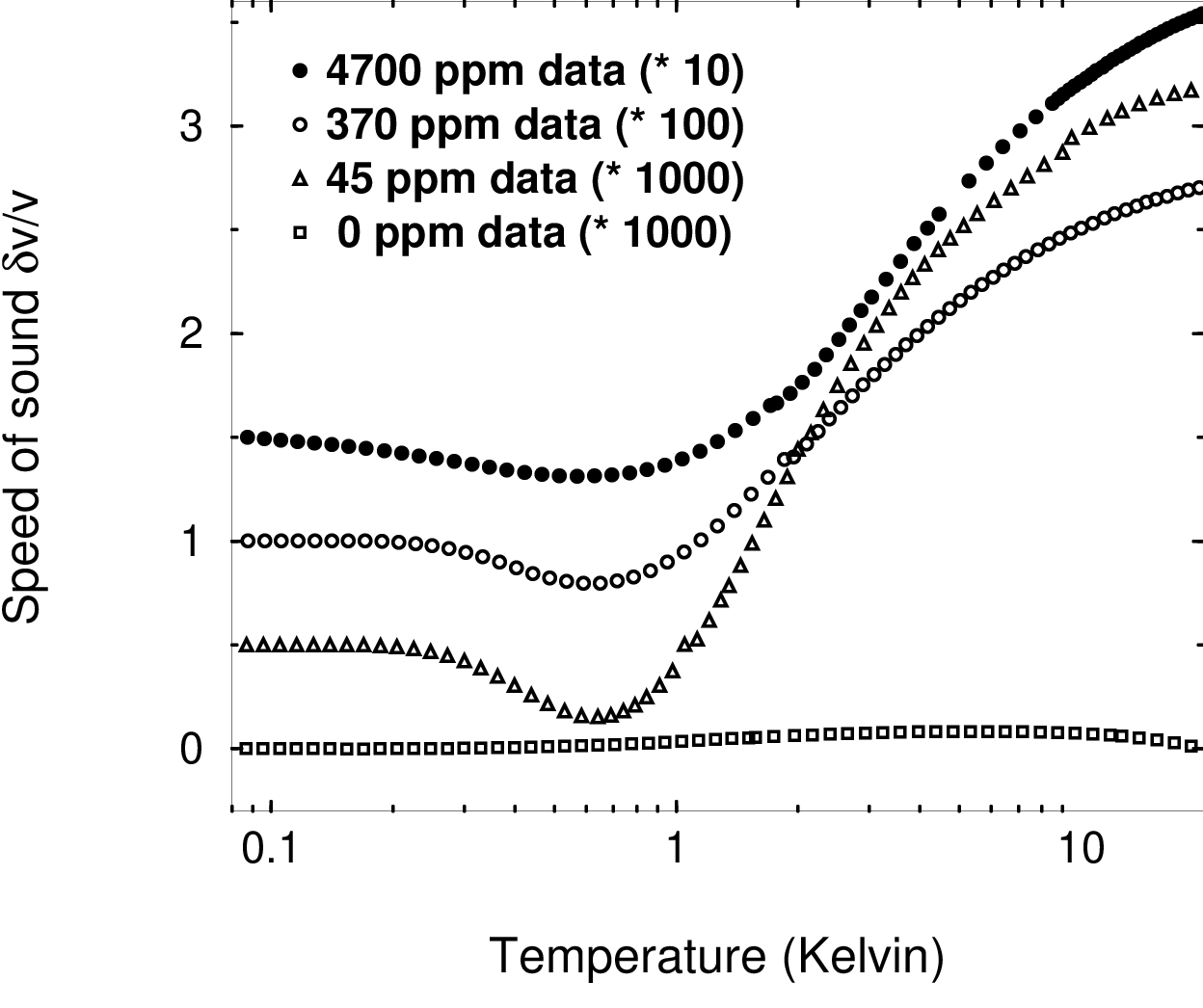,width=8cm,angle=270}} 
\caption{\label{dvv_Daten} Relative change of sound velocity $\delta v/v$ 
     for four KCl samples with different concentrations of CN$^-$ dopants 
     are plotted versus the temperature. For convenience, the data are 
     multiplied by the factors of 10. Note the vertical shift by steps 
     of 0.5 with respect to the somewhat arbitrary origin at zero 
     temperature.} 
\end{figure}   

Figure \ref{Q_Daten} shows the internal friction as a function 
of temperature, cyanide concentration ranging from 0 to 4700 ppm.
The data consist of several peaks. The dominant feature 
occurs at about 450 mK in the 45ppm-data and moves to about 350 mK 
in the 370ppm-data; in the strongly doped sample (4700ppm) it results 
in a broad shoulder beyond the lowest temperatures investigated. At 
higher $T$ there are additional peaks with much weaker intensity, 
which may well result from strongly coupled defect pairs \cite{chr}. 
The internal friction of the undoped sample is significantly smaller, 
although still remarkably high.  (Even after a check for purity, there
is still some unexpected damping in the ``pure'' KCl.  The expected
background internal friction value for pure crystals using this technique
is less than 3~$\times~10^{-6}$ below 1~K, as measured in quartz 
\cite{topp99}.) 

The relative change of the sound velocity is plotted in figure 
\ref{dvv_Daten}. The temperature dependence of the data may be 
decomposed in two features. First, there is a contribution proportional 
to $\tanh(E/2kT)$ with an energy $E$ of a few K, that is characteristic 
for quantum defects and shows the well-known $1/T$-dependence at 
higher temperature. Second, the minimum at $T=0.7$~K indicates a
relaxation process. 
  
Though the minimum broadens with increasing doping, the three samples 
show quite a similar behavior, and the change in sound velocity is 
roughly linear in the impurity concentration.

\section{Comparison of theory and experiment} 

We focus on the data of KCL with 45 ppm CN$^-$ since our theory is
correct for low doping only, i.e. for impurities without
interaction. With increasing doping, however, their dipolar
interactions are no longer small. The collective dynamics results in
the broad internal friction spectrum shown in figure \ref{Q_Daten} at
higher concentrations; the sharp features indicate strongly coupled
impurity pairs.

The theoretical expressions for symmetric tunnel impurities involve the
tunnel energy $\Delta$ and the damping rate $\Gamma_0$ (\ref{e46});
taking also finite asymmetry energies as in Sect. \ref{asymmetry} into
account, the Gaussian width
$\sigma$ provides one more parameter. Both the rate $\Gamma_0$ and the
prefactors of the internal friction and the change of sound velocity
(\ref{e58},\ref{e60}) depend on the deformation potentials $\gamma_\alpha$,
the sound velocities $v_\alpha$, and the mass density of the host crystal
$\varrho$. The latter quantities are well known; since the longitudinal
sound velocity is significantly larger than the transverse one, we have
$v_t^{-5}\gg v_l^{-5}$ and retain transverse sound waves only, with
$v_t=1.7$~km/s and $\varrho=1.989$ g/cm$^3$. (We neglect the weak
dependence of the sound velocity on the propagation direction.)

The tunnel
energy $\Delta/k_B$ of CN in KCl has been measured by various techniques,
with different experimental results; paraelectric resonance: 1.87~K
\cite{Hol79}; excitation of optical vibrations: 1.73~K \cite{Luet74};
specific heat: 1.6~K \cite{Per69}.
The relatively large discrepancy between these values may be due
to experimental uncertainties and different parameters of the samples,
in particular impurity concentration. From our fits to the 45 ppm data we
obtained a value of 1.55~K, which is in reasonable agreement with the above
results. Regarding the asymmetry energy, we use $\sigma/k_B=0.0145$~K for
the width of the Gaussian distribution. Finally, the transverse deformation
potential $\gamma_t$ constitutes the most important parameter, appearing
both in the prefactors and relaxation rates. From our fits we obtain the
value $\gamma_t=0.192$~eV which is of the same order of magnitude as
the values obtained for OH impurities in KCl (1~eV) \cite{Roberts}, in NaCl
(1.0~eV) \cite{Burst} and OD in NaCl (0.34~eV) \cite{Burst}. 
 
\begin{figure}
\centering{\epsfig{file=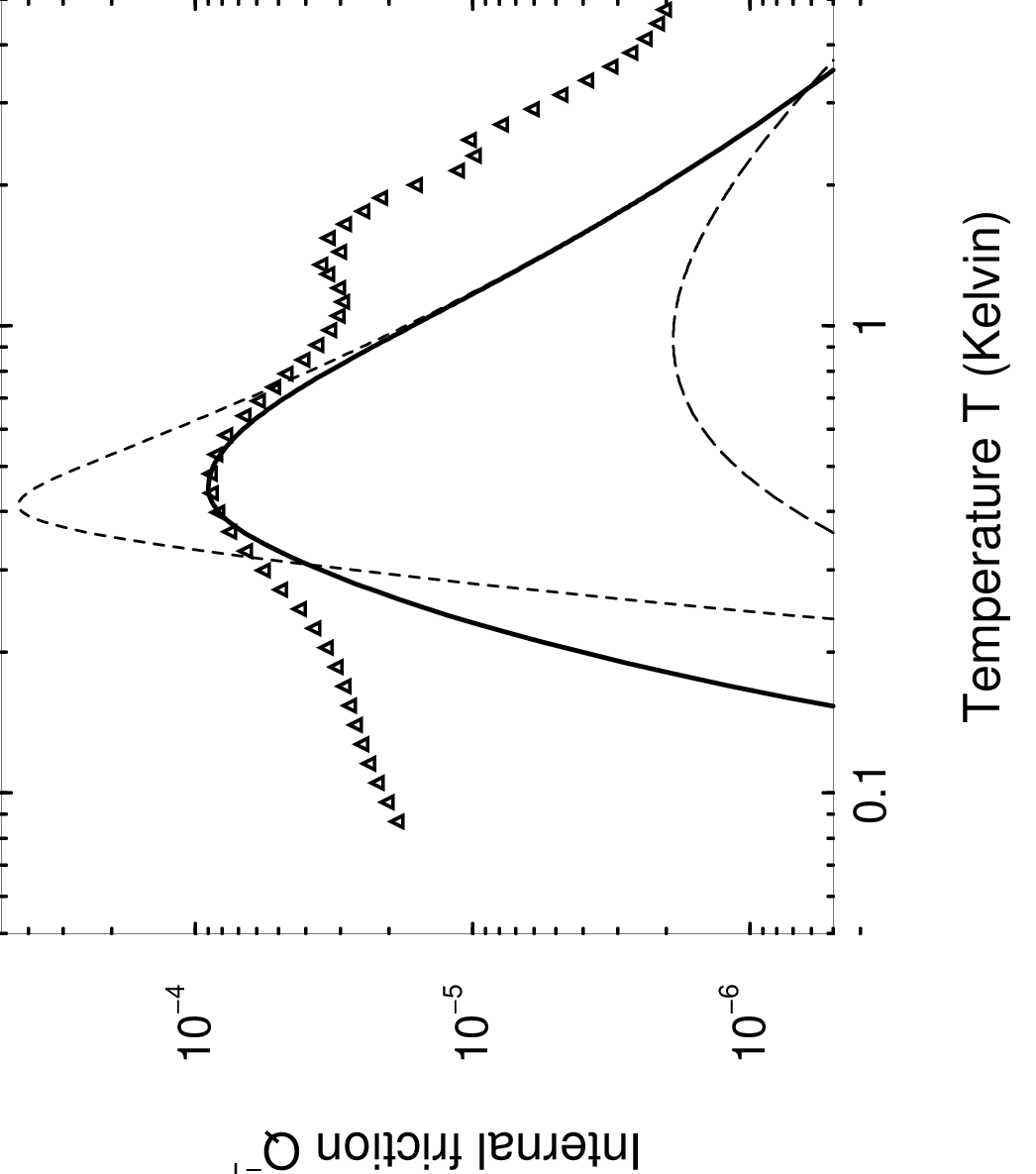,width=8cm,angle=270}}
\caption{\label{intfric}The [111]-defect theory is fitted at the data 
       of the internal friction of the KCl: 45ppm CN$^-$ sample. The
       full line is a fit including asymmetry effects and the short-dashed 
       line without these. The long-dashed line shows a plot of the
internal friction of a two-level-system where we set $(\epsilon/E)=1$
whereas all other parameters was choosen the same as in the former fit.} 
\end{figure} 

First we discuss the internal friction as shown in figure
\ref{intfric}. The triangles are the background corrected data of the
45 ppm sample, i.e., we have subtracted from the original data the 
values for the undoped sample. The full line is a fit including 
asymmetry effects (\ref{chiasy2}) where we used the parameters
given in Table 1. 

The temperature where the maximum occurs depends strongly on the
tunneling amplitude $\Delta$ and the fraction $\omega/\Gamma_0$;
the maximum value of $Q^{-1}$ varies with the prefactor and the
asymmetry width as far as the mean splitting is bigger than the
frequency $\bar{\eta}=\sigma^2/\Delta\ge\omega$.

The relaxation peak arises where the external frequency is comparable
to the slow rate $\Gamma_2$, and  $T_{max}$ is determined by the
relation $\Gamma_2(T_{max})=\omega$. Thus the exponential decrease of
$\Gamma_2$ is essential for the existence of the relaxation
maximum.

Note that this mechanism does not exist for two-level tunneling systems,
even with large asymmetries, since their relaxation rate tends
towards a constant $\Gamma_0$. As a consequence, for $\omega\ll\Gamma_0$ 
the relaxation maximum is suppressed by a factor $\omega/\Gamma_0$, and
the internal friction of a two-level system is by two orders of magnitude
smaller than that obtained for the [111]-impurity model. Note that the
parameters gathered in Table 1 result in a rate constant
$\Gamma_0=1.35\times 10^9$~sec$^{-1}$ which is indeed much larger than the
frequency $omega=5.3\times 10^5$~sec$^{-1}$.
For systems with $\eta>\omega$ the peak is suppressed by $\omega/\eta$ 
(see eq.(\ref{plo2})) which means that the peak contribution stems from
systems with small asymmetries. Therefore the peak in the internal
friction data determines only a combination of the prefactor and the
asymmetry width. 

Our simple model accounts well for position and height of the peak and
for its shape close to the maximum. Yet it strongly underestimates the
wings at temperatures well above and below $T_{\rm max}$. As to the excess 
spectral weight observed at low temperatures, a more realistic
distribution of asymmetry energies would probably give a better
agreement than the Gaussian used in (\ref{chiasy2}). The additional
peak at about 2~K may well be due to strongly coupled pairs. 

\begin{figure}
\centering{\epsfig{file=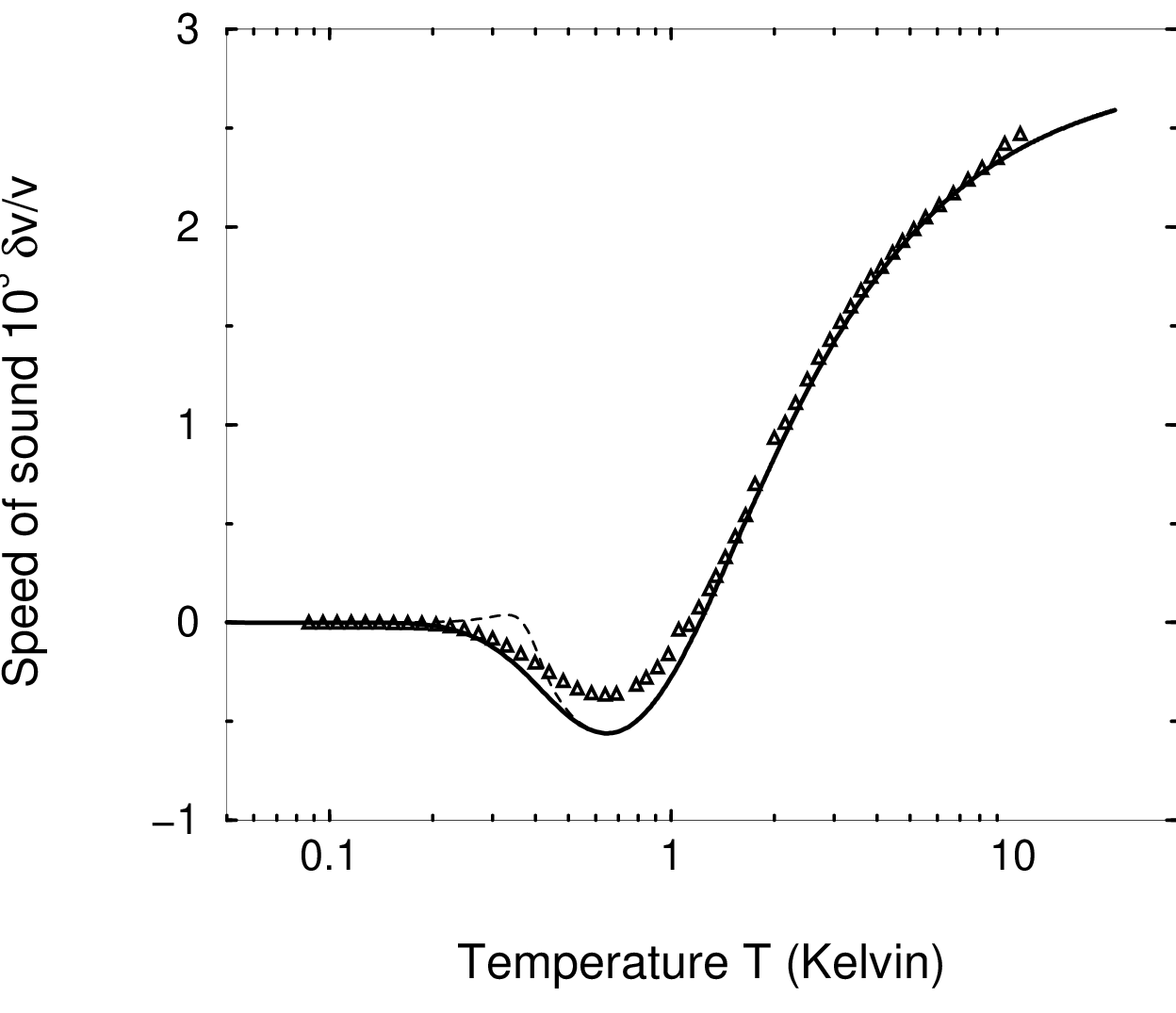,width=8cm,angle=270}}
\caption{\label{soundvel}The [111]-defect theory is fitted at the data 
         of the relative change in speed of sound of the KCl: 45ppm CN$^-$ 
         sample. The full line is a fit including asymmetry effects 
         and the dashed line without these.} 
\end{figure} 

Now we turn to the relative change of sound velocity as shown in figure
\ref{soundvel}. The triangles are data for the 45 ppm sample. The dashed
line is a fit without asymmetry, given by (\ref{e60}), and the full line
with asymmetry (\ref{chiasy1}). The asymmetry
distribution with width $\sigma$ merely smears out the temperature
dependence. Thus the fit of Fig. \ref{soundvel} mainly depends on the
tunneling amplitude and the prefactor but it hardly varies with the
relaxation rate and the asymmetry width.

\begin{table}[t]
\begin{center}
\begin{tabular}{|c|c|c|c|c|c|c|}
\hline
   $\omega/2\pi$ (sec$^{-1}$) & $n$ (cm$^{-3}$) & $\varrho$ (gcm$^{-3}$) &
   $v_t$ (km/s) & $\Delta/k_B$ (K) & $\gamma_t$ (eV) & $\sigma/k_B$ (K) 
\cr
\hline
   84352 &  $1.7\times10^{17}$ & 1.989 & 1.7 & 1.55 & 0.192 & 0.0145
\cr
\hline
\end{tabular}
\end{center}
\caption{\label{tabelle1} Parameters used for the fits of
Figs. \ref{intfric} and \ref{soundvel}} 
\end{table}

Regarding both Figs. \ref{intfric} and \ref{soundvel}, our model
fits remarkably well the absolute values of $\delta v/v$ and $Q^{-1}$, 
the hump in the sound velocity, and the position of the relaxation
peak in the internal friction. The transverse deformation potential
is the most relevant parameter that determines both the absolute
values of $\delta v/v$ and $Q^{-1}$, and the relaxation maximum
of the latter; we find the value $\gamma_t=0.192$~eV. (Note the
factor 2 in our definition of the phonon coupling (\ref{e12}).)

The width of the relaxation peak of $Q^{-1}$ and the low-temperature
wing indicate the relevance of the asymmetry energies (\ref{asymmetrie}).
The poor quality of our fit at low $T$ may well be due to the choice
of a Gaussian distribution for the asymmetries. Nonetheless, our
results confirm that for the 45 ppm sample, the mean asymmetry is by
two orders of magnitude smaller than the tunnel energy, which is
consistent with the model assumptions.

The present theory accounts fairly well for available data on the
relaxation behavior of [111]-impurities in alkali halides investigated
with $T_{2g}$-modes, in
particular with respect to the temperature dependence. Yet the
multiexpenontial decay with the two rates $\Gamma_1$ and $\Gamma_2$
gives rise to quite an intricate relaxation spectrum; thus an
experimental investigation of the frequency dependence would seem
most promising. A simliar behaviour was also found for an $E_g$-mode
\cite{ByerSack} contradicting the simplest model of an
[111]-defect. Even asymmetries can not explain this feature in a
simple way.
Another open question concerns the thermal conductivity \cite{Nara}. The
relaxation feature obtained in the present work should significantly
contribute to thermal resistivity at low T. Still, the spectrum of the
defects obtained from the data \cite{Nara} suggests a more complicated
elastic response spectrum and may well require large asymmetry energies.

Another interesting point would be to extend the present work to the case of
strong doping; this is certainly not an easy taks, given the serious
difficulties encountered already in the treatment of interacting
two-state systems (see, e.g., Ref. \cite{Kue2000}.) Note, however,
that our sound velocity data in Fig. \ref{dvv_Daten} depend, roughly
speaking, linearly on the impurity concentration. Thus it would seem
that they do not fullfil the strong-coupling criterion of
\cite{Wur97}, i.e., the dipolar interactions have not yet destroyed
the coherent tunnel motion. 

\section{Summary} 

We have investigated the relaxation of an impurity ion in alkali halides
arising from the coupling to elastic waves. We briefly summarize
the main results.

(i) The various elastic and inelastic phonon-mediated transition
between the eight quantum states give rise to an intricate
temperature and frequency dependence of the relaxation contributions
to the internal friction and the sound velocity (see Fig. \ref{spectrum}.)
Unlike two-state tunneling systems, [111]-impurities 
show two relaxation rates. At low $T$, the smaller rate decreases
exponentially, $\Gamma_2=e^{-2\Delta/kT}\Gamma_0$, whereas the larger one
tends towards a constant, $\Gamma_1=\Gamma_0$. This corresponds
to a multiexponential decay of the time-dependent response and correlation
functions (\ref{e15}) and (\ref{e50}).

(ii) A most particular relaxation behavior arises for external frequencies
$\omega$ that are smaller than the constant rate, $\omega<\Gamma_0$. Then
the relaxation maximum occurs at a temperature where the smaller rate
$\Gamma_2$ is equal to the external frequency $\omega$. This maximum is the
more relevant as at low $T$ the spectral weight of the slow contribution
exceeds by far that of the faster rate $\Gamma_1$.

(iii) Comparison with recent data on KCl:CN proves the relevance
of this relaxation mechanism. The exponentially decreasing rate
$\Gamma_2$ explains the large amplitude of the relaxation peak
shown by the external friction $Q^{-1}$ as a function of
temperature at $\nu=84$ kHz. By the same token, our model provides
a good description for the hump in the sound velocity. 

(iv) The prefactors of $\delta v/v$ and $Q^{-1}$ and the relaxation
rates are related by the values for the deformation potential $\gamma$,
sound velocity $v$, and mass density $\rho$. By taking $\gamma$ as a free
parameter, we obtain for reasonable values of $\gamma$ (compared with
simliar materials) satisfying fits for $\delta v/v$ and $Q^{-1}$,
involving both absolute values and the temperature dependencies. 

(v)  The phenomena mentioned in (iii) cannot be explained in terms of
relaxation of
corresponding two-level systems. For the latter, the low-temperature
rate tends towards a constant and thus may exceed $\omega$ at any $T$,
whereas the small rate $\Gamma_2$ of a [111]-impurity inevitably meets
$\omega$ at some $T$ and gives rise to a relaxation peak. Though
certain aspects of the thermal and dielectric properties of such
doped crystals are described by a ensemble of two-state systems,
such a model fails in view of the acoustic properties, due to the
multiexponential decay of the elastic response function (\ref{e15}).  

{\bf ACKNOWLEDGEMENTS}
We would like to express a special thank you to Prof. Robert Pohl for
his guidance on the experiments as well as many stimulating
discussions and helpful comments on this manuscript.
We also wish to thank C. Enss for stimulating the theoretical
investigation and H. Horner, R. K\"uhn and B. Thimmel for many helpful
discussions.  
Additionaly P. Nalbach wants to thank the DFG which supported the work
within the DFG-project HO 766/5-3 "Wechselwirkende 
Tunnelsysteme in Gl\"asern und Kristallen bei tiefen Temperaturen".

\begin{appendix} 
\section{Geometric factor of the ESS} 

Time evolution of the phonon heat bath is given by the lattice 
response function $(t\ge0)$ 
\begin{equation} 
\label{app2} 
\Gamma(t) = \sum_\alpha \frac{\gamma_\alpha^2}{\hbar^2} 
\overline{ 
\langle [\epsilon_{ij}^{\alpha}(t),\epsilon_{ij}^{\alpha}]\rangle}, 
\end{equation} 
where $i\ne j$ and the bar indicates the average over crystal axes. 
The entries of the damping matrix (\ref{e26}) are determined 
by the coupled phonon spectrum; in the limit of long wavelengths one has 
\begin{equation} 
\label{app4} 
\Gamma''(\omega) = \frac{\pi}{2} \sum_{{\bf k} \alpha}f^{(ij)}_{{\bf
k}\alpha} \gamma_\alpha^2 
{k^2 \over m \hbar\omega_{{\bf k}\alpha}} 
[ \delta(\omega-\omega_{{\bf k}\alpha}) 
- \delta(\omega+\omega_{{\bf k}\alpha}) ], 
\end{equation} 
where 
\begin{equation} 
\label{app6} 
f^{(ij)}_{{\bf k}\alpha} =  \left(e_i^{\alpha}({\bf k})\hat k_j 
+e_j^{\alpha}({\bf k})\hat k_i \right)^2  
\end{equation} 
accounts for the orientations of wave and polarizations vectors with 
respect to the crystal axes. 
For an isotropic phonon density of states, the sum over phonon modes 
becomes 
\begin{equation} 
\label{app8} 
\frac1 V \sum_{{\bf k},\alpha} f^{(ij)}_{{\bf k}\alpha}(...) \longrightarrow 
{1\over 2\pi^2}\sum_{\alpha}f_\alpha \int_0^K \mbox{d}k k^2 (...) 
\end{equation} 
where we have for each polarization defined the average value 
\begin{equation} 
\label{app10} 
f_\alpha = {1\over 4\pi}\int \mbox{d}\Omega f^{(ij)}_{{\bf k \alpha}} . 
\end{equation} 
In the Debye model with $\omega_{{\bf k}\alpha}=v_\alpha k$, the damping 
spectrum finally reads as 
\begin{equation} 
\label{app12} 
\Gamma''(\omega) = \frac\pi 2 \sum_{{\bf k}\alpha}f_\alpha 
{\gamma_\alpha^2 \over v_\alpha^5} 
{\omega^3 \over 2\pi^2 \rho \hbar}. 
\end{equation}
Since an elastic wave couples via each of the quadrupole operators to
the defect according to (\ref{e12}), the geometric factor $h_{{\bf
k}\alpha}$ of the internal friction and the 
relative change of the sound velocity of an elastic wave with
propagation vector $\bf k$ and polarisation vector ${\bf e}^\alpha$ gets  
\begin{equation} 
\label{app16}
h_{{\bf k}\alpha} = f^{(12)}_{{\bf k \alpha}} + f^{(13)}_{{\bf k
\alpha}} + f^{(23)}_{{\bf k \alpha}}
\end{equation}
The factors $f_\alpha$ are easily evaluated after expressing the unit 
vectors $\hat{\bf k}={\bf k}/k$ and ${\bf e}$ in polar coordinates 
$\theta$ and $\phi$. Putting $\hat{\bf k}=(\sin\theta \sin\phi, 
\sin\theta \cos\phi,\cos\theta)$, one finds for the longitudinal case 
${\bf e}^1=\hat{\bf k}$ the factor $f_l=4/15$. 

Regarding the transverse modes, we obtain $f_{t_2}=8/45$ for ${\bf e}^2 
=(\cos\phi,-\sin\phi,0)$ and $f_{t_3}=2/9$ for the remaining 
polarization ${\bf e}^3$. Though these values depend on the choice of 
${\bf e}^2$ and ${\bf e}^3$, their average $f_t=(1/2)(f_{t_2}+f_{t_3}) 
=1/5$ is independent of the basis. 
Defining moreover the mean value of the three polarisation directions, 
$f=(1/3)(f_{l}+2f_{t})$, we finally have 
\begin{equation}\label{app14} 
f_l={4\over15},\hskip0.5cm f_t={1\over5},\hskip0.5cm f={2\over9}. 
\end{equation} 
Thus the quantity (\ref{app6}) for longitudinal modes is by a factor 
$f_l/f_t=4/3$ larger than for the transverse ones. In the average, 
propagation and polarisation directions are uncorrelated, resulting 
in 
\begin{equation} f = \left( \langle \hat k_i^2\rangle\langle e_j^2\rangle + 
\langle \hat k_j^2\rangle\langle e_i^2\rangle\right) = 2 (1/3)^2
\end{equation}

\end{appendix}

\end{document}